\documentclass[aps,physrev,reprint,superscriptaddress]{revtex4-2}

\usepackage{graphicx}
\usepackage{dcolumn}
\usepackage{bm}
\usepackage[mathlines]{lineno}
\usepackage{upgreek}    
\usepackage{siunitx}    
\usepackage{braket}     
\usepackage{amsmath}
\usepackage{amssymb}
\usepackage{hyperref} 

\begin{document}


\title{Coherent electronic Raman excitation of valley-orbit split states of phosphorus dopants in silicon}


\author{Adam Gindl}%
    \thanks{Contact author: adam.gindl@matfyz.cuni.cz}
    \affiliation{Faculty of Mathematics and Physics, Charles University, Ke Karlovu 3, 12116 Prague 2, Czech Republic}
\author{Martin \v{C}mel}
    \affiliation{Faculty of Mathematics and Physics, Charles University, Ke Karlovu 3, 12116 Prague 2, Czech Republic}
\author{Franti\v{s}ek Troj\'{a}nek}
    \affiliation{Faculty of Mathematics and Physics, Charles University, Ke Karlovu 3, 12116 Prague 2, Czech Republic}
\author{Petr Mal\'{y}}
    \affiliation{Faculty of Mathematics and Physics, Charles University, Ke Karlovu 3, 12116 Prague 2, Czech Republic}
\author{Zbyn\v{e}k \v{S}ob\'{a}\v{n}}
    \affiliation{Institute of Physics, Czech Academy of Sciences, Cukrovarnick\'{a} 10, 162 00 Prague 6, Czech Republic}
\author{Alexandr Po\v{s}ta}
    \affiliation{Faculty of Electrical Engineering, Czech Technical University in Prague, Technick\'{a} 2, 166 27 Prague, Czech Republic}
\author{Martin Koz\'{a}k}
    \affiliation{Faculty of Mathematics and Physics, Charles University, Ke Karlovu 3, 12116 Prague 2, Czech Republic}

\begin{abstract}
In this study, we demonstrate coherent optical excitation of the electronic Raman transition between the $1s\left(A_1\right)$ and $1s\left(E\right)$ split states of phosphorus donor in crystalline silicon. The dynamics of the generated wavepacket is characterized in the time domain using a degenerate pump-probe technique with mid-infrared femtosecond pulses via transient polarization anisotropy of the probe pulse. In addition, we study the role of resonantly excited carriers, and we show that the amplitude and coherence time of the electronic wavepacket depend on the pre-excited carrier density. Further, we demonstrate that under certain conditions, the Raman-type excitation changes to displacive impulsive excitation, which allows us to address the Raman-forbidden transition between $1s\left(A_1\right)$ and $1s\left(T_1\right)$.
\end{abstract}

\maketitle

\section{Introduction}

The structure of quantum energy levels in matter can be studied using Raman spectroscopy, utilizing the inelastically scattered photons. Since the discovery of the Raman effect in 1928 \cite{Raman1928_1, Raman1928_2, Raman1928_3, Raman1928_4}, Raman spectroscopy has evolved into one of the fundamental classes of optical methods offering relatively easy access to chemical and structural properties of the studied medium \cite{Das2011}. The most widely used technique, referred to as vibrational Raman spectroscopy, operates by exciting vibrational modes in the material through the inelastic scattering of incident photons. This approach enables the study of vibrations of specific chemical bonds in molecules \cite{Galván2013, Liu2017} or phonon modes in solids \cite{Weinstein1975, Kash1985}. Although many other techniques exploit the Raman effect to study different features of excitations in the material medium, such as rotational modes of molecules \cite{Ozer2024, Satija2020}, magnons in magnetic systems \cite{Kuroe2008, Rigitano2020}, or even electronic excitations in semiconductors \cite{Klein1983}.

The doping of semiconductors represents an indispensable part of modern electronics. However, it also introduces new possible electron transitions of considerable importance for the function of electronic components. Thus, it is of utmost importance to be able to describe these in detail. The first theoretical study of dopant states in the energy band gap of a host semiconductor used effective mass theory with a hydrogenic impurity potential $U\left(\vec{r}\right)=-\frac{1}{4\pi\varepsilon_0}\frac{e^2}{r\varepsilon}$, where $\varepsilon_0$ is vacuum permittivity, $e$ is the elementary charge, $r$ is radial coordinate with respect to impurity atom and $\varepsilon$ is the static dielectric constant of the semiconductor \cite{Luttinger1955, Kohn1955}. For shallow donors in crystals with multiple energy-degenerate conduction band minima at $\vec{k}\neq0$, known as \textit{valleys}, the effective mass approximation predicts a degenerated donor ground state below the conduction band. However, early measurements of photoexcitation spectra of \textit{n}-Si \cite{Aggarwal1964, Aggarwal1965, Toyotomi1975}, \textit{n}-Ge \cite{Fisher1962, Reuszer1964}, \textit{n}-AlSb \cite{Ahlburn1968} and Hall measurements in \textit{n}-GaP \cite{Toyama1968}  revealed significant differences in binding energies of donor bound states, referred to as \textit{chemical shift}, and splitting of its ground state into fine multiplet. Extensive theoretical studies were conducted to accurately predict the spectra of newly introduced impurities resulting from \textit{n}-doping. Early methods, which employed different variational approaches and central-cell corrections to the hydrogenic effective mass theory, partially addressed the problem of chemical shift \cite{Weinreich1959, Muller1964, Csavinszky1965}, but did not explain the splitting of the ground state. The creation of multiplet, known as \textit{valley-orbit splitting}, resulting from the multivalley structure of \textit{n}-doped semiconductor was eventually revealed by considering intervalley mixing \cite{Baldereschi1970, Pantelides1974, Shindo1976, Altarelli1977, Pantelides1978, Klymenko2015, Revin2021}, which was neglected in all previous studies using a single-valley approach.

There is high demand for multivalley semiconductors in the evolving field of valleytronics, which exploits the electron quantum number associated with the wavevector $\vec{k}$ of the populated valley for information storage and processing, similarly to the electron electric charge used in electronics. Most commonly used valleytronic materials are monolayers of transition-metal dichalcogenides \cite{Cao2012, Mak2018, Langer2018}. However, various approaches using bulk diamond and silicon \cite{Isberg2013, Hammersberg2014, Suntornwipat2021, Gindl2025}, or monolayers and multilayers of group-IV monochalcogenides \cite{Hanakata2016, Lin2018, Hashmi2025, Pan2025} are also being developed nowadays. Therefore, a precise and accessible experimental technique is necessary to study the influence of bound electron transitions between valley-orbit split donor states in \textit{n}-doped multivalley semiconductors.

Raman scattering by electronic excitations in semiconductors (\textit{electronic Raman scattering}) is a widely used technique that offers insight into these excitations and their properties \cite{Klein1983}. This method was already used to observe excitations of free electrons, plasmons, and coupled plasmon-phonon excitations \cite{Mooradian1966, Mooradian1967}, or various non-plasmonic electron-phonon \cite{Cerdeira1973_1, Cerdeira1973_2}, donor-phonon \cite{Dean1970}, and exciton-phonon excitations \cite{Levinson1973}.

In this work, we report on ultrafast coherent electronic Raman scattering by excitation of bound electron transition between valley-orbit split $1s\left(A_1\right)$ and $1s\left(E\right)$ states of phosphorus donors in silicon crystal (indirect band gap $E_g = \SI{1,12}{\electronvolt}$) due to interaction with an ultrashort nonresonant strong-field infrared pulse. Previously, only incoherent electronic Raman spectra from scattering by $1s$ phosphorus donor states were observed with the strongest line at approximately $\SI{13,1}{\milli\electronvolt}$, corresponding to $1s\left(A_1\right)\rightarrow1s\left(E\right)$ transition \cite{Wright1967}, measured for different concentrations of phosphorus donors in silicon \cite{Jain1976, Stavrias2019}.

\section{Valley-orbit splitting of donor ground state}

In this section, we present the derivation of the multivalley effective mass equation describing the origin of valley-orbit split donor levels in \textit{n}-doped multivalley crystals, specifically in phosphorus-doped silicon \cite{Luttinger1955, Kohn1955, Kohn1957, Baldereschi1970, Pantelides1974, Bassani1974, Shindo1976, Pantelides1978}. Such system is described by Schr\"{o}dinger equation
\begin{equation} \label{eq1:PerturbedSchrodingerEquation}
    \left[ \left( \frac{p^2}{2m_0} + V_0\left(\vec{r}\right) \right) + U\left(\vec{r}\right) \right] \psi\left(\vec{r}\right) = E \psi\left(\vec{r}\right),
\end{equation} where $\vec{p}$ is momentum operator, $m_0$ is free electron mass, $V_0\left(\vec{r}\right)$ is periodic potential of perfect crystal, $U\left(\vec{r}\right)$ is impurity potential, $\psi\left(\vec{r}\right)$ is impurity wavefunction and $E$ represents energy of impurity states. The term in round brackets corresponds to the Hamiltonian of a perfect crystal described by the equation
\begin{equation} \label{eq2:UnperturbedSchrodingerEquation}
    \left( \frac{p^2}{2m_0} + V_0\left(\vec{r}\right) \right) \psi_{n,\vec{k}}^0\left(\vec{r}\right) = E_{n,\vec{k}} \psi_{n,\vec{k}}^0\left(\vec{r}\right),
\end{equation} where $n$ is band index, $\psi_{n,\vec{k}}^0\left(\vec{r}\right) = u_{n,\vec{k}}\left(\vec{r}\right) \exp{\left(i\vec{k}\cdot\vec{r}\right)}$ are electron Bloch functions, $u_{n,\vec{k}}\left(\vec{r}\right)$ is periodic function with periodicity of crystal lattice and $E_{n,\vec{k}}$ represents energy band structure of unperturbed crystal. The perturbation is then introduced by an impurity, whose wavefunction can be expanded in terms of the Bloch functions
\begin{equation} \label{eq3:WavefunctionExpansion}
    \psi\left(\vec{r}\right) = \sum_{n,\vec{k}\in BZ} F_{n,\vec{k}} \psi_{n,\vec{k}}^0\left(\vec{r}\right),
\end{equation} where $F_{n,\vec{k}}$ are envelope functions. For a shallow donor, it is sufficient to consider only the conduction band in the previous expansion. Thus, the band index will be dropped in the following derivation \cite{Pantelides1978}.

Using expansion (\ref{eq3:WavefunctionExpansion}) and equation (\ref{eq2:UnperturbedSchrodingerEquation}) in impurity Schr\"{o}dinger equation (\ref{eq1:PerturbedSchrodingerEquation}), multiplying on the left by complex conjugate Bloch function, integrating over all space, and considering orthonormality of Bloch functions, we obtain
\begin{equation} \label{eq4:AlmostEME}
\begin{split}
    E_{\vec{k}} F_{\vec{k}} &+ \sum_{\vec{k}'\in BZ} F_{\vec{k}'} \int u_{\vec{k}}^*\left(\vec{r}\right) e^{-i\vec{k}\cdot\vec{r}} U\left(\vec{r}\right) \\
    &\times u_{\vec{k}'}\left(\vec{r}\right) e^{i\vec{k}'\cdot\vec{r}} d\vec{r} = E F_{\vec{k}}.
\end{split}    
\end{equation} Due to the periodicity of the functions $u_{\vec{k}}\left(\vec{r}\right)$ corresponding to the periodicity of the crystal lattice, we can expand their product into plane waves
\begin{equation} \label{eq5:UProductExpansion}
    u_{\vec{k}}^*\left(\vec{r}\right) u_{\vec{k}'}\left(\vec{r}\right) = \sum_p C_{\vec{k},\vec{k}'}^p e^{i\vec{K}_p \cdot \vec{r}},
\end{equation} where $C_{\vec{k},\vec{k}'}^p$ is expansion coefficient and $\vec{K}_p$ is vector of reciprocal lattice of perfect crystal. By substituting (\ref{eq5:UProductExpansion}) in (\ref{eq4:AlmostEME}), we obtain a term representing the Fourier transform of the impurity potential $\tilde{U}\left(\vec{k} - \vec{k}' - \vec{K}_p \right)$, which leads to the general form of the effective mass equation represented in momentum space
\begin{equation} \label{eq6:GeneralEME}
    E_{\vec{k}} F_{\vec{k}} + \sum_{\vec{k}'\in BZ} \sum_p C_{\vec{k},\vec{k}'}^p F_{\vec{k}'} \tilde{U}\left(\vec{k} - \vec{k}' - \vec{K}_p \right) = E F_{\vec{k}}.
\end{equation}

In a case of silicon crystal, which has six valleys close to the X points at the Brillouin zone edge in [100], [010] and [001] crystallographic direction, the envelope function from (\ref{eq3:WavefunctionExpansion}) can be written as superposition of functions $F_{\vec{k}}^{\left(i\right)}$ localized around $i$-th valley
\begin{equation} \label{eq7:EnvelopeFuncExpansion}
    F_{\vec{k}} = \sum_{i=1}^6 \alpha_i F_{\vec{k}}^{\left(i\right)},
\end{equation} where expansion coefficients $\alpha_i$ can be determined from the symmetry of irreducible representations of the cubic tetrahedral group $T_d$ related to the phosphorus impurity \cite{Kohn1955, Pantelides1974, Jain1976, Kohn1957}. Consequently, we can divide the Brillouin zone into six subzones $\Omega_i$, each being centered at a position of one valley $\vec{k}_{i}$ \cite{Bassani1974}. The effective mass equation (\ref{eq6:GeneralEME}) then splits into six coupled multivalley equations
\begin{equation} \label{eq8:MultivalleyEME}
\begin{split}
    E_{\vec{k}}^{\left(i\right)} \alpha_i F_{\vec{k}}^{\left(i\right)} &+ \sum_{j=1}^6 \sum_{\vec{k}'\in\Omega_j} \sum_p C_{\vec{k},\vec{k}'}^p \alpha_j \\
    &\times F_{\vec{k}'}^{\left(j\right)} \tilde{U}\left(\vec{k} - \vec{k}' - \vec{K}_p \right) = E \alpha_i F_{\vec{k}}^{\left(i\right)},
\end{split}
\end{equation} where $\vec{k}\in\Omega_i$ for $i=1,\dots,6$ and $E_{\vec{k}}^{\left(i\right)}$ is second-order term of conduction band energy expansion in ($\vec{k}-\vec{k}_i$). For the valley at $\left(k_0,0,0\right)$, this corresponds to
\begin{equation} \label{eq9:ConductionBandEnergy}
    E_{\vec{k}}^{\left(i\right)} = \frac{\hbar^2}{2m_\parallel} \left(k_x - k_0\right)^2 + \frac{\hbar^2}{2m_\perp} \left(k_y^2 + k_z^2\right),
\end{equation} where $m_\parallel = 0.92 m_0$ is longitudinal electron effective mass, $m_\perp = 0.19 m_0$ is transverse electron effective mass \cite{Hensel1965}.

Early approaches omitted proper dielectric screening of impurity potential, resulting in negligible values of intervalley ($i\neq j$) and Umklapp ($p\neq0$) terms compared to intravalley terms ($i=j$) in equations (\ref{eq8:MultivalleyEME}). This results in six uncoupled equivalent single-valley equations
\begin{equation} \label{eq10:SingleValleyEME}
    E_{\vec{k}}^{\left(i\right)} \alpha_i F_{\vec{k}}^{\left(i\right)} + \sum_{\vec{k}'\in\Omega_i} \alpha_i F_{\vec{k}'}^{\left(i\right)} \tilde{U}\left(\vec{k} - \vec{k}' \right) = E \alpha_i F_{\vec{k}}^{\left(i\right)},
\end{equation} where we considered $C_{\vec{k},\vec{k}'}^0 = 1$, leading to a six-fold degenerate $1s$ ground state of phosphorus dopant \cite{Luttinger1955, Kohn1955, Weinreich1959, Muller1964, Csavinszky1965}. The hydrogenic impurity potential screened by a proper wavevector-dependent dielectric function $U\left(\vec{r}\right)=-\frac{1}{4\pi\varepsilon_0}\frac{e^2}{r\varepsilon\left(\vec{k}\right)}$ \cite{Nara1965, Vinsome1971} can be used as a reasonable approximation for real impurity potential in a case of isocoric impurities, when the dopant and host have the same number of core electrons, such as phosphorus donor in silicon \cite{Pantelides1974}. Then, the resulting intervalley terms are no longer negligible, equations (\ref{eq8:MultivalleyEME}) are coupled, and their symmetry is reduced, in comparison to the symmetry of uncoupled single-valley equations (\ref{eq10:SingleValleyEME}), to the symmetry of the studied system \cite{Baldereschi1970}. Therefore, the original six-fold degeneracy of the ground state is partially lifted. Also, including Umklapp terms introduces a valley-dependent renormalization factor for a properly screened impurity potential, which further improves the chemical shift values of valley-orbit split $1s$ states \cite{Altarelli1977} in comparison to the experimental measurements of phosphorus donor photoexcitation spectra \cite{Aggarwal1964, Aggarwal1965, Jagannath1981, Mayur1993}.

The six-fold degenerate $1s$ ground state of phosphorus donor in silicon introduced in single-valley effective mass theory is thus split into three levels due to valley-orbit splitting: singlet $1s\left(A_1\right)$, doublet $1s\left(E\right)$, and triplet $1s\left(T_1\right)$ with binding energies of $\SI{45,59}{\milli\electronvolt}$, $\SI{32,58}{\milli\electronvolt}$ and $\SI{33,89}{\milli\electronvolt}$ respectively \cite{Jagannath1981}. The singlet state $1s\left(A_1\right)$ thus corresponds to the donor electron ground state. Notation $A_1$, $E$, and $T_1$ of individual states represents their symmetry and degeneracy, corresponding to irreducible representations of the tetrahedral group $T_d$ of phosphorus donors in silicon \cite{Kohn1955}. Latest works calculated positions of $1s$ energy levels theoretically utilizing more advanced techniques, such as diagonalization of the donor Hamiltonian in the basis of unperturbed crystal Bloch functions \cite{Wellard2005}, Burt-Foreman envelope function representation in effective mass theory \cite{Klymenko2015}, or ab initio approach using density functional theory \cite{Smith2017}.

\section{Coherent electronic Raman scattering}

Valley-orbit transitions between $1s$ phosphorus states in silicon can be induced by coherent electronic Raman scattering. We will derive the expression for the differential Stokes-Raman scattering cross-section \cite{Klein1983, Jain1976, Colwell1972, Yoon2005}. First, we rewrite the impurity wavefunction
\begin{equation} \label{eq11:WavefunctionRaman}
    \psi^{\left(\nu\right)}\left(\vec{r}\right) = \sum_{j=1}^{6} \alpha_j^{\left(\nu\right)} F_{j}\left(\vec{r}\right) u_{\vec{k_j}}\left(\vec{r}\right) e^{i \vec{k_j} \cdot \vec{r}},
\end{equation} where $\nu = 1, \dots, 6$ represents singlet state $1s\left(A_1\right)$, doublet state $1s\left(E\right)$, and triplet state $1s\left(T_1\right)$, $F_{j}\left(\vec{r}\right) = \sum_{\vec{k} \in BZ} F_{\vec{k}}^{\left(j\right)} e^{i\left(\vec{k}-\vec{k_j}\right) \cdot \vec{r}}$ is the transformation of the envelope function from momentum space, used in equation (\ref{eq3:WavefunctionExpansion}), to coordinate space and $\alpha_j^{\left(\nu\right)}$ are expansion coefficients already introduced in equation (\ref{eq7:EnvelopeFuncExpansion}) corresponding to irreducible representation $\nu$. The Hamiltonian describing the interaction between incident radiation and a single donor electron in the crystal is given by
\begin{equation} \label{eq12:HamiltonianRaman}
\begin{split}
    H &= \frac{1}{2m_0}\left[\vec{p} - e\vec{A}\left(\vec{r},t\right) \right]^2 + V_0\left(\vec{r}\right) + U\left(\vec{r}\right) \\
    &= H^{\left(0\right)} + H^{\left(1\right)} + H^{\left(2\right)},
\end{split}
\end{equation} where $\vec{A}\left(\vec{r},t\right)$ is vector potential of electromagnetic field. Valley-orbit splitting is given by the term unperturbed by radiation
\begin{equation} \label{eq13:UnperturbedHamiltonianRaman}
    H^{\left(0\right)} = \left( \frac{p^2}{2m_0} + V_0\left(\vec{r}\right) \right) + U\left(\vec{r}\right),
\end{equation} which was used in the derivation of the multivalley effective mass equation (\ref{eq8:MultivalleyEME}). The light-matter interaction is then represented by the linear term
\begin{equation} \label{eq14:LinearHamiltonianRaman}
    H^{\left(1\right)} = -\frac{e}{m_0} \vec{p} \cdot \vec{A}\left(\vec{r},t\right),
\end{equation} and the nonlinear term
\begin{equation} \label{eq15:NonlinearHamiltonianRaman}
    H^{\left(2\right)} = \frac{e^2}{2m_0} A^2\left(\vec{r},t\right).
\end{equation}

To calculate the differential Stokes-Raman scattering cross-section for the transition between the initial $\ket{0}$ and the final donor state $\ket{f}$, we first apply the dipole approximation and the effective mass approximation. We use time-dependent perturbation theory, where the $H^{\left(2\right)}$ term is used in first-order and $H^{\left(1\right)}$ term in second-order. The differential cross-section is then given by
\begin{equation} \label{eq16:CrossSectionStart}
    \frac{d\sigma}{d\Omega}\left(0\rightarrow f\right) = \frac{\omega_R}{\omega_L} r_0^2 \left|\vec{e_L} \cdot \vec{e_R} \Braket{f | 0} + \frac{1}{m_0} \mathcal{M}^{\left(2\right)} \right|^2,
\end{equation} where $\omega_L$ ($\omega_R$) are angular frequencies of laser (Raman) photons, $\vec{e_L}$ ($\vec{e_R}$) are unit polarization vectors of laser (Raman) photons, $r_0 = \frac{1}{4\pi\varepsilon_0}\frac{e^2}{m_0c^2}$ is classical electron radius, $\varepsilon_0$ is permittivity of free space and $c$ is speed of light in vacuum. The second-order term is given by
\begin{equation} \label{eq17:SecondOrderTerm1}
\begin{split}
    \mathcal{M}^{\left(2\right)} &= \sum_m \left[ \frac{\Bra{f} \vec{p} \cdot \vec{e_L} \Ket{m} \Bra{m} \vec{p} \cdot \vec{e_R} \Ket{0}}{E_0-E_m-\hbar\omega_R} \right. \\
    &\left. + \frac{\Bra{f} \vec{p} \cdot \vec{e_R} \Ket{m} \Bra{m} \vec{p} \cdot \vec{e_L} \Ket{0}}{E_0-E_m+\hbar\omega_L} \right],
\end{split}
\end{equation} where $\Ket{m}$ is intermediate state during Raman scattering with energy $E_m$ and $E_0$ is energy of the initial state.

To derive the cross-section of valley-orbit transitions, we use a two-band model, where the valley in the conduction band at position $\vec{k_j}$ corresponds to energy $E_c$, the maximum of the valence band corresponds to energy $E_v$, and the band gap is $E_g = E_c -E_v$. Three different Raman transitions can occur between donor levels: (a) valley-orbit transition between ground state and higher $1s$ state via virtual intermediate state in the valence band, (b) excitation from ground state to excited hydrogenic state (e.g. $2s$ state) via virtual intermediate state in the valence band, or (c) excitation from ground state to excited even-parity state (in or outside $1s$ levels) via virtual intermediate odd-parity hydrogenic state (e.g. $2p$ state). However, it was shown that processes (b) and (c) are negligible in comparison to process (a) \cite{Colwell1972}. 

In process (a), the energy of the intermediate state $E_m = E_v$, and we can approximate the energy of the donor electron ground state as $E_0 \approx E_c$. The second-order term (\ref{eq17:SecondOrderTerm1}) is then given by
\begin{equation} \label{eq18:SecondOrderTerm2}
    \mathcal{M}^{\left(2\right)} = \sum_{\alpha,\beta=1}^3 e_L^\alpha e_R^\beta \left[ \frac{\Bra{f} p_\alpha p_\beta \ket{0}}{E_g-\hbar\omega_R} + \frac{\Bra{f} p_\beta p_\alpha \ket{0}}{E_g + \hbar\omega_L} \right].
\end{equation} Wavefunctions of initial and final donor states are given by equation (\ref{eq11:WavefunctionRaman}), where envelope functions are slowly varying in position space. Therefore, we can assume that momentum operators do not act on them. For the valley-orbit transition within the $1s$ manifold, the envelope function overlap integral $\Braket{F_j^f | F_j^0} = 1$. Also, the energy difference between valley-orbit split states is much smaller than the energy band gap $E_g \gg \hbar\left(\omega_L - \omega_R\right)$. The second-order term is then written as
\begin{equation} \label{eq19:SecondOrderTerm3}
\begin{split}
    \mathcal{M}^{\left(2\right)} &= \frac{2E_g}{E_g^2 - \left(\hbar\omega_L\right)^2} \sum_{\alpha,\beta=1}^3 e_L^\alpha e_R^\beta \sum_{j=1}^6 \alpha_j^{\left(0\right)} \alpha_j^{\left(f\right)\ast}\\
    &\times\Bra{u_{\vec{k_j}}} \left(\vec{p}+\hbar\vec{k_j}\right)_\alpha \left(\vec{p}+\hbar\vec{k_j}\right)_\beta \Ket{u_{\vec{k_j}}}.
\end{split}
\end{equation} We rewrite the bra-ket term from the previous equation in terms of the reciprocal effective mass tensor using $\vec{k}\cdot\vec{p}$ perturbation theory
\begin{equation} \label{eq20:MomentumOperatorsToEffectiveMassTensor}
\begin{split}
    \Bra{u_{\vec{k_j}}} \left(\vec{p}+\hbar\vec{k_j}\right)_\alpha \left(\vec{p}+\hbar\vec{k_j}\right)_\beta \Ket{u_{\vec{k_j}}} = \\
    = \frac{m_0}{2}E_g\left[ \left( \frac{m_0}{m^\ast} \right)_{\alpha\beta}^{\left(j\right)} - \delta_{\alpha\beta} \right],
\end{split}
\end{equation} where the dimensionless reciprocal effective mass tensor can be written for the case of axial symmetry as
\begin{equation} \label{eq21:EffectiveMassTensorAxialSymmetry}
     \mathbf{\left( \frac{m_0}{m^\ast} \right)^{\left(j\right)}} = \frac{m_0}{m_\perp}\mathbf{I} + \left( \frac{m_0}{m_\parallel} - \frac{m_0}{m_\perp} \right) \vec{e_j} \otimes \vec{e_j},
\end{equation} where $\mathbf{I}$ is identity matrix and $\vec{e_j}$ is unit vector along the direction of valley $\vec{k_j}$. For the valley-orbit transition, where $\Ket{0}\neq\Ket{f}$, differential Stokes-Raman scattering cross-section is given by
\begin{equation} \label{eq22:CrossSectionFinal}
\begin{split}
    \frac{d\sigma}{d\Omega}\left(0\rightarrow f \right) &= \frac{\omega_R}{\omega_L} r_0^2 R_{12}^2 \left( \frac{m_0}{m_\parallel} - \frac{m_0}{m_\perp} \right)^2 \\
    &\times \left| \vec{e_L} \cdot \left[ \sum_{j=1}^6 \alpha_j^{\left(0\right)} \alpha_j^{\left(f\right)\ast} \vec{e_j} \otimes \vec{e_j} \right] \cdot \vec{e_R} \right|^2,
\end{split}    
\end{equation} where $R_{12} = \frac{E_g^2}{E_g^2 - \left(\hbar\omega_L\right)^2}$ is resonance enhancement factor.

Selection rules for valley-orbit Raman transition are then decided by values of numerical coefficients $\alpha_j^{\left(\nu\right)}$ and by the orientation of laser and Raman photons polarization with respect to the position of the individual valleys in momentum space. In case of the silicon crystal, numerical coefficients given by standard group theoretical technique are \cite{Kohn1955, Kohn1957}
\begin{equation} \label{eq23:NumericalCoefficients}
    \begin{aligned}
        \vec{\alpha}^{\left(A_1\right)} &= \frac{1}{\sqrt{6}} \left(1,1,1,1,1,1\right), \\
        \vec{\alpha}^{\left(E-1\right)} &= \frac{1}{2} \left(1,1,-1,-1,0,0\right), \\
        \vec{\alpha}^{\left(E-2\right)} &= \frac{1}{2} \left(1,1,0,0,-1,-1\right), \\
        \vec{\alpha}^{\left(T_1-1\right)} &= \frac{1}{\sqrt{2}} \left(1,-1,0,0,0,0\right), \\
        \vec{\alpha}^{\left(T_1-2\right)} &= \frac{1}{\sqrt{2}} \left(0,0,1,-1,0,0\right), \\
        \vec{\alpha}^{\left(T_1-3\right)} &= \frac{1}{\sqrt{2}} \left(0,0,0,0,1,-1\right).
    \end{aligned}
\end{equation} The coefficients can be derived alternatively using Frobenius reciprocity theorem, however, all the coefficients remain the same, except for the $\vec{\alpha}^{\left(E-2\right)} = \frac{1}{2\sqrt{3}} \left(1,1,1,1,-2,-2\right)$ \cite{Jain1976, Ramdas1981}. In both cases, the sum over $j$ in square brackets in equation (\ref{eq22:CrossSectionFinal}) is nonzero only for the transition from donor ground state $1s\left(A_1\right)$ to excited doublet state $1s\left(E\right)$, and transition to triplet state $1s\left(T_1\right)$ is forbidden.

\section{Experimental details}

\begin{figure}[h!]
\includegraphics[width=\linewidth]{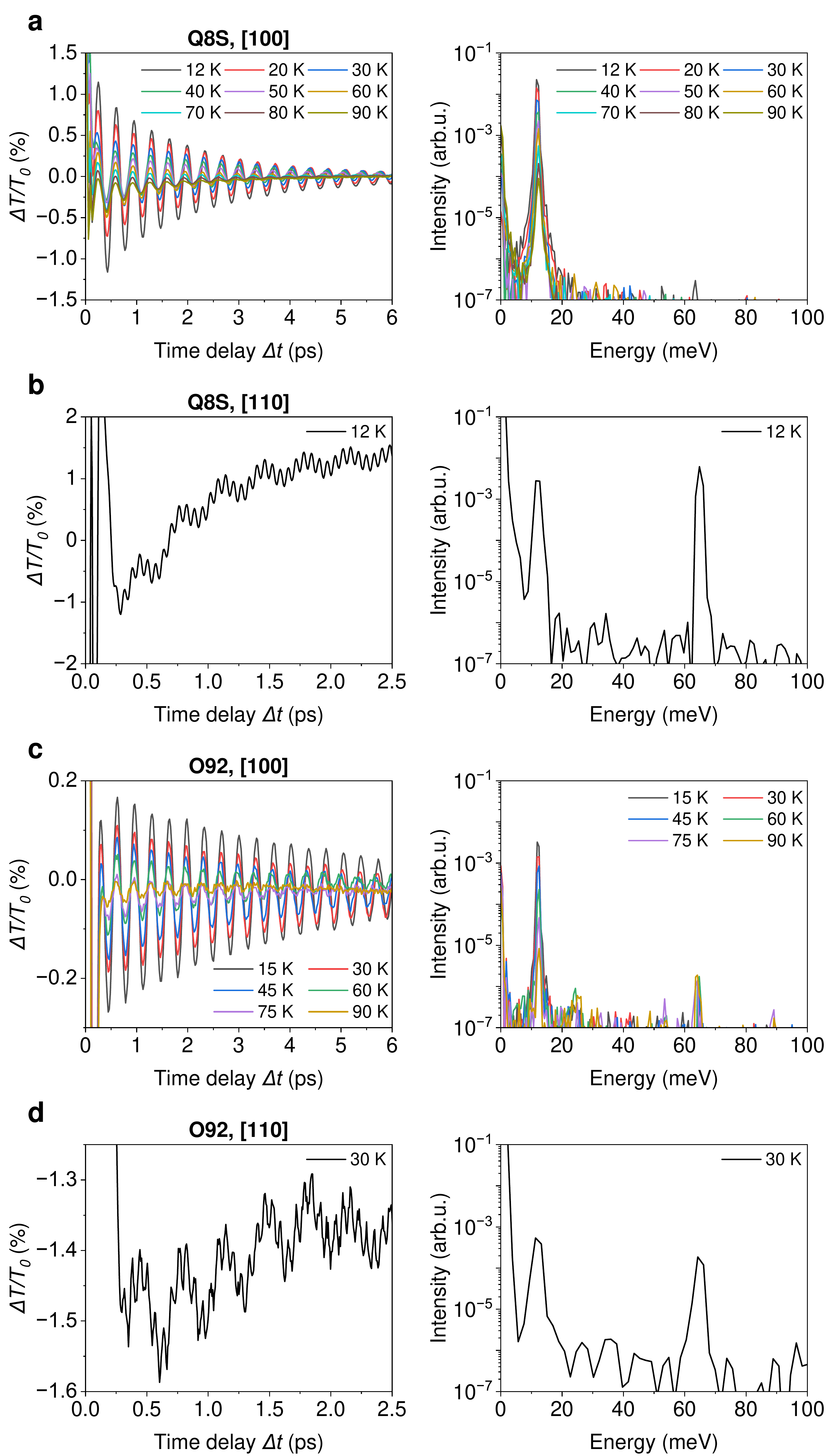}
\centering
\caption{Measured periodic coherent oscillations of probe pulse transient transmittance polarization anisotropy $\Delta T/T_0$ of phosphorus-doped silicon samples (left) and corresponding Fourier-transformed spectrum (right) at different temperatures. Data were measured for two samples with different dopant concentration (O92 - lower dopant concentration, Q8S - higher dopant concentration) and two orientations of crystal with respect to pump pulse linear polarization: (\textbf{a}) Q8S sample with pump pulse linearly polarized along $\left[100\right]$ direction, (\textbf{b}) Q8S sample with pump pulse linearly polarized along $\left[110\right]$ direction, (\textbf{c}) O92 sample with pump pulse linearly polarized along $\left[100\right]$ direction, and (\textbf{d}) O92 sample with pump pulse linearly polarized along $\left[110\right]$ direction. Pump pulse has peak intensity of $I_0 = \SI{2,65e11}{\watt\per\square\centi\meter}$. Pre-excitation pulse was not used. Fourier-transformed spectra are displayed on a logarithmic scale.}
\label{fig1}
\end{figure}

In our measurements, coherent excitation induced by the first strong-field infrared femtosecond pulse, corresponding to the generation of a coherent wavepacket as a linear superposition of bound electron eigenstates, is probed with a second time-delayed infrared pulse, which investigates the evolution of the wavepacket. Both pulses are linearly polarized, and the time delay between them is controlled with an optical delay line. Two components of the probe pulse field, orthogonal and parallel to the pump pulse polarization, are detected separately on two InGaAs photodiodes. Then, the difference signal between these two is measured using a lock-in amplifier. This approach offers femtosecond temporal resolution, allowing us to study ultrafast dynamics of excited bound electrons in phosphorus donors. Coherent excitation is observed in the form of periodic oscillations of the polarization anisotropy of transient transmittance of the crystal (left side of Fig. \ref{fig1}) with oscillation frequency corresponding to the binding energy difference between $1s\left(A_1\right)$ and  $1s\left(E\right)$ valley-orbit split eigenstates (right side of Fig. \ref{fig1}).

Both infrared pulses used in pump-probe measurements have approximately normal incidence on the sample. Pulses have photon energy of $\SI{0,62}{\electronvolt}$, duration of $\SI{40}{\femto\second}$ and repetition rate of $\SI{25}{\kilo\hertz}$. We used a variable neutral density filter to adjust the peak intensity of the pump pulse on the sample. Probe pulse has linear polarization rotated by $\SI{45}{\degree}$ with respect to the pump pulse polarization. Pulses were generated in a non-collinear optical parametric amplifier with a subsequent difference frequency generation setup pumped by an ytterbium-based femtosecond laser system \cite{Kozák2021}. The fundamental output of the ytterbium-based femtosecond laser was used in some of the measurements as a pre-excitation pulse with a photon energy of $\SI{1,2}{\electronvolt}$ and a pulse duration of $\SI{170}{\femto\second}$, to control the density of excited carriers in the sample. To ensure that the pre-excited carriers are already relaxed to band minima and their distribution functions can be described as a thermalized distributions at the moment when the pump pulse arrives at the sample, the pre-excitation pulse is incident on the sample $\SI{100}{\pico\second}$ before the pump pulse.

Coherent electronic oscillations were observed using transmission geometry pump-probe measurements in two monocrystalline silicon samples with different phosphorus concentration purchased from MicroChemicals GmbH: higher-doped sample Q8S (data shown in Fig. \ref{fig1}a,b) with resistivity quoted by manufacturer $\rho = \qtyrange[range-phrase = -,range-units = single]{0,02}{0,05}{\ohm\centi\metre}$ corresponding to dopant concentration $N_D = \qtyrange[range-phrase = -,range-units = single]{2,63e17}{1,63e18}{\per\cubic\centi\metre}$, and lower-doped sample O92 (data shown in Fig. \ref{fig1}c,d) with resistivity quoted by manufacturer $\rho = \qtyrange[range-phrase = -,range-units = single]{0,1}{1}{\ohm\centi\metre}$ corresponding to dopant concentration $N_D = \qtyrange[range-phrase = -,range-units = single]{4,95e15}{8,35e16}{\per\cubic\centi\metre}$ \cite{Masetti1983, Sze1968}. Both samples have a thickness $d = \sisetup{separate-uncertainty}\qty{300\pm25}{\micro\meter}$ and a surface orthogonal to the $\left[001\right]$ crystallographic direction, coated with an antireflection layer to suppress undesirable pump pulse reflection from the back surface of the sample, which would lead to its additional time-shifted excitation. The \SI{340}{\nano\metre} SiO$_2$ coating layers were deposited using electron-beam evaporation and atomic layer deposition (ALD) processes. The final thickness of the SiO$_2$ layers was verified by ellipsometry and additionally confirmed by profilometric measurement of the physical step height. Each sample of \textit{n}-Si was placed in a closed-cycle helium cryostat, allowing us to control the crystal temperature in the range from $\SI{12}{\kelvin}$ to $\SI{295}{\kelvin}$.

\section{Results}

Two orientations of the crystalline sample were used with respect to the linear polarization of the pump pulse. In the first case, the linear polarization of the pump pulse is oriented along the $\left[100\right]$ crystallographic direction, while in the second case, it is oriented along the $\left[110\right]$ direction. Due to electron effective mass tensor anisotropy of silicon crystal, introduced in equation (\ref{eq9:ConductionBandEnergy}), pump pulse polarized along $\left[100\right]$ direction (data shown in Fig. \ref{fig1}a,c) accelerates electrons in valleys along $\left[100\right]$ and $\left[010\right]$ axes to different energies, generating slow, weak, exponentially decaying signal from anisotropic electron population, known as \textit{valley polarization}, given by energy-dependent intervalley electron-phonon scattering \cite{Isberg2013, Gindl2025, Jacoboni1983}. This type of signal is not observed for pump pulses linearly polarized along the $\left[110\right]$ direction (data shown in Fig. \ref{fig1}b,d), as they interact with electrons in both types of valleys equally. However, another faster oscillations are observed as a consequence of excitation of coherent optical phonons with an energy of $\SI{65}{\milli\electronvolt}$ (see spectra in Fig. \ref{fig1}) \cite{Kulda1994}, which were negligible for $\left[100\right]$ polarization, due to the symmetry of the phononic Raman tensor of silicon \cite{Shinohara2010_1, Ishioka2006, Hase2003, Shinohara2010_2}. Additionally, there is an exponentially decaying background from the carrier response \cite{Zeiger1992}. For both polarizations, the fast signal at $\Delta t = \SI{0}{\pico\second}$ originates from the coherent interaction of the pump and probe pulses, due to their spatial and temporal overlap.

We studied properties of coherent valley-orbit transition $1s\left(A_1\right)\rightarrow1s\left(E\right)$ as a function of crystal temperature, peak intensity of the pump pulse, and density of pre-excited carriers. Measured data are shown in Fig. \ref{fig1}. The data are processed as follows. Firstly, the exponentially decaying background from the carrier response in $\left[110\right]$ polarization data (Fig. \ref{fig1}b,d) was eliminated using FFT-based spectral filtering. The resulting data of transient transmittance polarization anisotropy $\Delta T/T_0$ were then fitted using a sum of two exponentially decaying cosine functions in the case of $\left[110\right]$ orientation
\begin{equation} \label{eq24:ExpDecayCosPhonon}
\begin{split}
    \Delta T/T_0 = A_e\cos{\left(2\pi f_e\Delta t+\phi_e\right)}e^{-\Delta t/t_{e}}+\\
    +A_p\cos{\left(2\pi f_p\Delta t+\phi_p\right)}e^{-\Delta t/t_{p}}+y_0,
\end{split}
\end{equation} where the first term corresponds to the oscillations of the electronic wavepacket and the second term corresponds to phononic oscillations. In the case of $\left[100\right]$ orientation, the fitting function is an exponentially decaying cosine function with an exponentially decaying background corresponding to the relaxation of valley polarization 
\begin{equation} \label{eq25:ExpDecayCosExpDecBack}
\begin{split}
    \Delta T/T_0 = &A_e\cos{\left(2\pi f_e\Delta t+\phi_e\right)}e^{-\Delta t/t_{e}}+\\
    &+Be^{-\Delta t/t_{rel}}+y_0.
\end{split}
\end{equation} In expressions (\ref{eq24:ExpDecayCosPhonon}) and (\ref{eq25:ExpDecayCosExpDecBack}), $A_e$ ($A_p$) is amplitude of coherent electronic (phononic) oscillations, $f_e=E_{tr}/h$ ($f_p$) is electronic (phononic) oscillation frequency, $E_{tr}$ is central valley-orbit transition energy, $h$ is Planck constant, $\Delta t$ is pump-probe time delay, $\phi_e$ ($\phi_p$) is initial phase of coherent electronic (phononic) oscillations, $t_{e}$ ($t_{p}$) is electronic (phononic) coherence time, $B$ is amplitude of valley polarization signal, $t_{rel}$ is the relaxation time of valley polarization, and $y_0$ is constant background.

\begin{figure}[t]
\includegraphics[width=\linewidth]{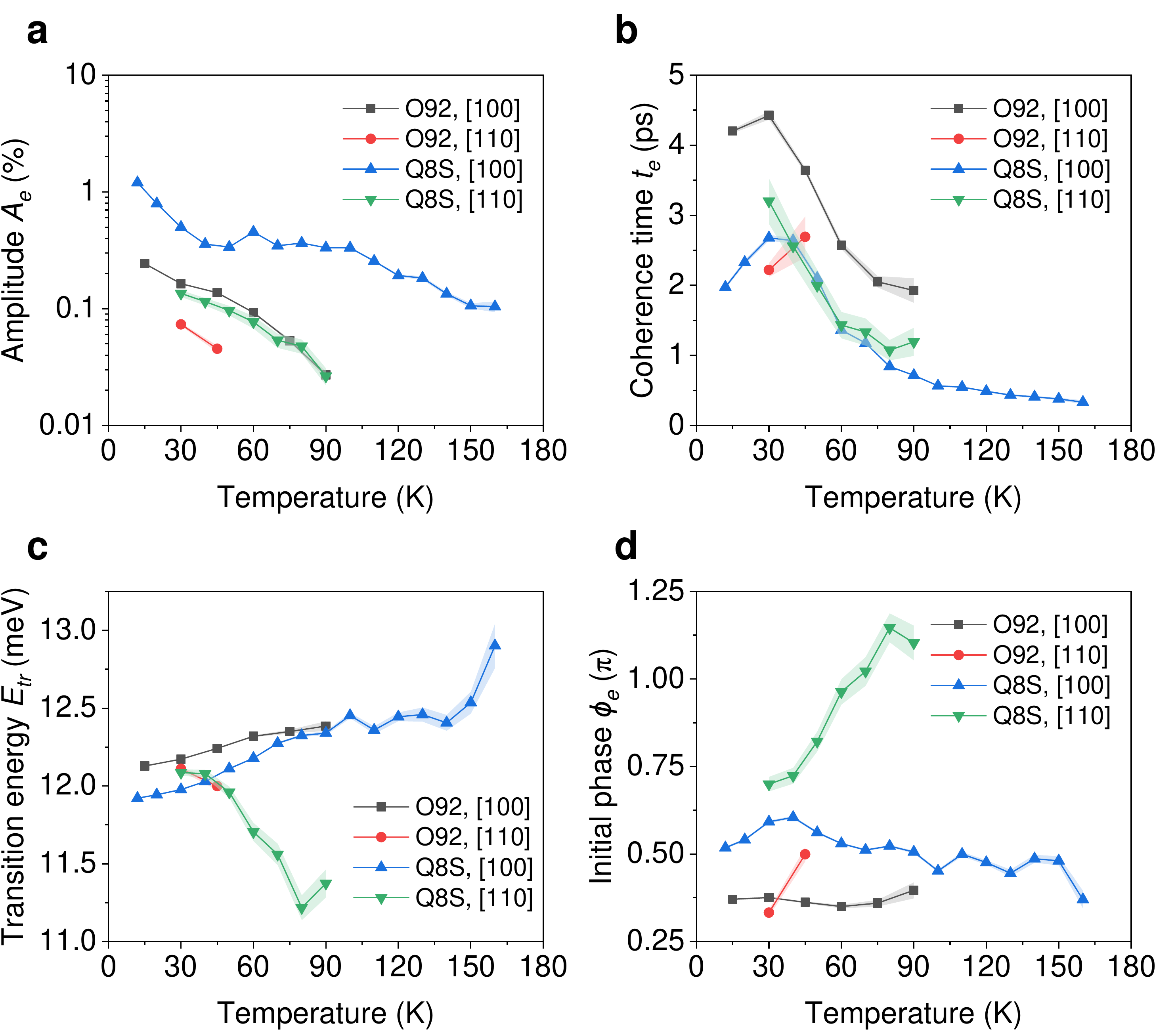}
\centering
\caption{Experimentally measured properties of coherent valley-orbit excitation in Q8S and O92 sample of phosphorus-doped silicon crystal as a function of crystal temperature. Data were measured for $\left[100\right]$ and $\left[110\right]$ orientation of pump pulse linear polarization in both samples. In Q8S measurements, pump pulse has peak intensity of $I_0 = \SI{1,66e11}{\watt\per\square\centi\meter}$ for its both polarizations. In O92 measurements, pump pulse has peak intensity of $I_0 = \SI{1,66e11}{\watt\per\square\centi\meter}$ for $\left[100\right]$ polarization and $I_0 = \SI{8,32e10}{\watt\per\square\centi\meter}$ for $\left[110\right]$ polarization. Measured parameters: (\textbf{a}) Amplitude of coherent oscillation in logarithmic scale, (\textbf{b}) coherence time, (\textbf{c}) central valley-orbit transition energy, and (\textbf{d})  initial phase of coherent oscillations. The shaded areas represent the standard deviation of parameters obtained by fitting experimental data of transient transmittance polarization anisotropy $\Delta T/T_0$ with the functions (\ref{eq24:ExpDecayCosPhonon}) and (\ref{eq25:ExpDecayCosExpDecBack}). Pre-excitation pulse was not used.}
\label{fig2}
\end{figure}

Acquired oscillation amplitude, coherence time, valley-orbit transition energy, and initial phase of coherent electronic oscillations as functions of crystal temperature are shown in Fig. \ref{fig2} for both samples and both orientations (for $\left[110\right]$ orientation of the O92 sample, only two points were obtained, as a consequence of rapidly vanishing electronic oscillations at higher temperatures). Both oscillation amplitude (Fig. \ref{fig2}a) and coherence time (Fig. \ref{fig2}b) are decreasing with increasing temperature. Firstly, a decrease in oscillation amplitude could be interpreted as the depletion of the $1s\left(A_1\right)$ ground state via thermal population of the $1s\left(E\right)$ excited state. Secondly, a shorter coherence time is given by the statistics of the same $1s\left(A_1\right)\rightarrow1s\left(E\right)$ incoherent thermal excitation, which is stronger with increasing temperature.  Additionally, the decoherence of the electron wave packet is faster for higher dopant concentrations, which is in agreement with previous measurements \cite{Jain1976}. This is a consequence of the increasing overlap between the wave functions of electrons on neighboring donors as the donor concentration increases. Therefore, the orbit of an electron spans more donor atoms, and its probability amplitude for being in the central cell of the donor decreases, which results in the decrease of valley-orbit splitting. Finally, electronic Raman scattering is considerably stronger for the Q8S sample due to a higher donor concentration \cite{Stavrias2019}, and for $\left[100\right]$ pump polarization due to a higher differential Stokes-Raman scattering cross-section, as determined by equation (\ref{eq22:CrossSectionFinal}).

\begin{figure}[t]
\includegraphics[width=\linewidth]{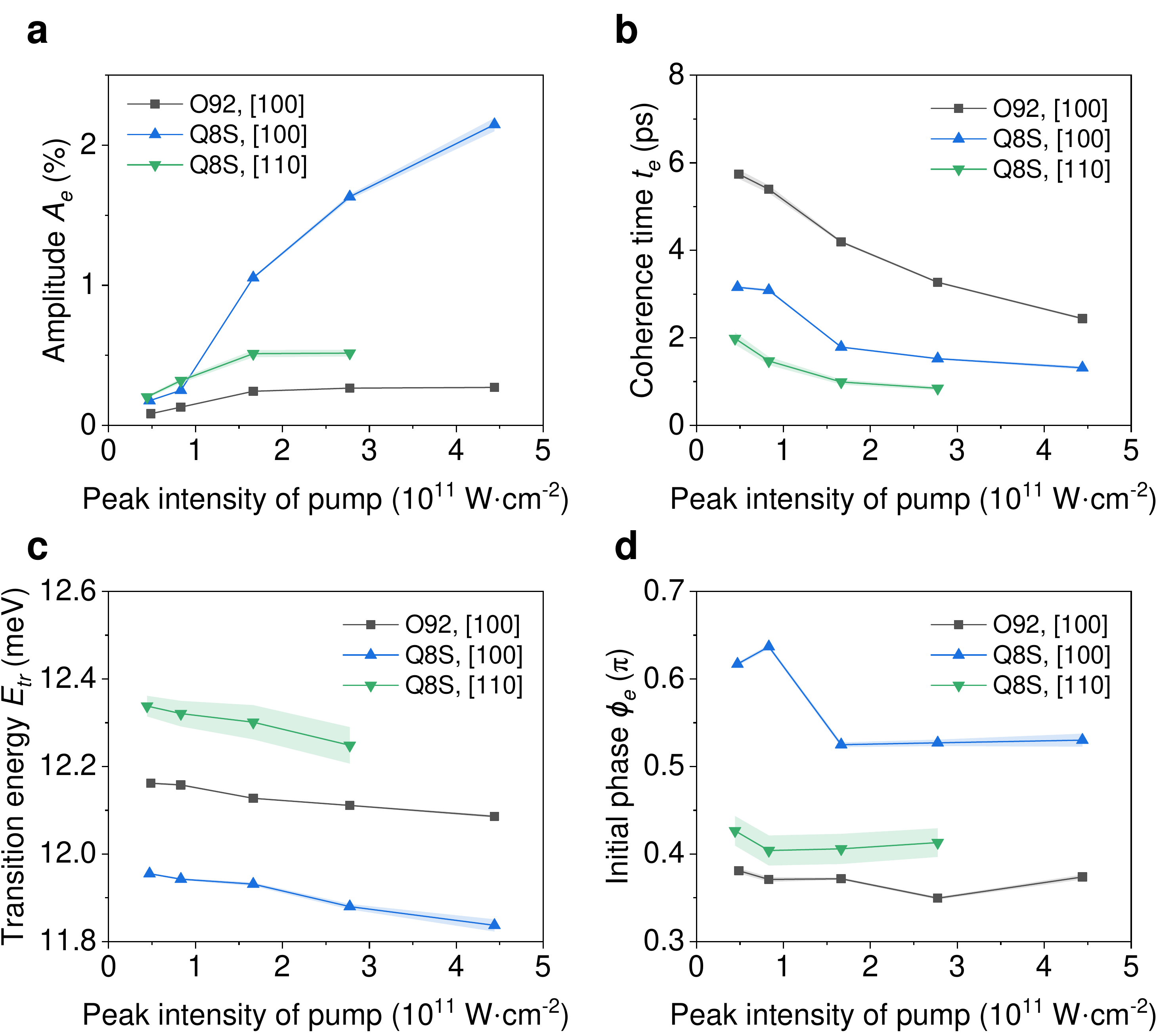}
\centering
\caption{Experimentally measured properties of coherent valley-orbit excitation in Q8S and O92 sample of phosphorus-doped silicon crystal at the temperature of 15 K as a function of peak intensity of pump pulse on the sample. Data were measured for both polarizations of the pump pulse in the Q8S sample, and for $\left[100\right]$ polarization in the O92 sample. Measured parameters: (\textbf{a}) Amplitude of coherent oscillation, (\textbf{b}) coherence time, (\textbf{c}) central valley-orbit transition energy, and (\textbf{d}) initial phase of coherent oscillations. The shaded areas represent the standard deviation of parameters obtained by fitting experimental data of transient transmittance polarization anisotropy $\Delta T/T_0$ with the functions (\ref{eq24:ExpDecayCosPhonon}) and (\ref{eq25:ExpDecayCosExpDecBack}). Pre-excitation pulse was not used.}
\label{fig3}
\end{figure}

The central valley-orbit transition energy (shown in Fig. \ref{fig2}c) is approximately the same for both samples and both orientations up to the crystal temperature of about $\SI{30}{\kelvin}$, corresponding to $1s\left(A_1\right)\rightarrow1s\left(E\right)$ Raman transition. Above this temperature, we observe a slowly increasing transition energy for the $\left[100\right]$ orientation in both samples, primarily due to changes in the silicon energy band gap as a function of temperature \cite{Bludau1974}, and rapidly decreasing transition energy for $\left[110\right]$ orientation towards the energy of Raman forbidden $1s\left(A_1\right)\rightarrow1s\left(T_1\right)$ transition. Qualitative change with temperature is observed also for the initial phase of coherent oscillations (Fig. \ref{fig2}d). The initial phase of electronic oscillations in the case of $\left[100\right]$ orientation is approximately constant at different temperatures, indicating a Raman transition. However, for $\left[110\right]$ orientation, values of initial phase are increasing with increasing temperature, indicating the occurrence of another type of excitation mechanism. Such behavior then corresponds to the situation, where in $\left[100\right]$ case, mainly $1s\left(A_1\right)\rightarrow1s\left(E\right)$ Raman transition is observed, but in $\left[110\right]$ case, $1s\left(A_1\right)\rightarrow1s\left(T_1\right)$ transition starts to predominate with increasing temperature, due to another nonlinear mechanism, despite being Raman-forbidden. Our hypothesis is that in this case, the excitation of the forbidden transition occurs due to the generation of electron-hole pairs via indirect dopant absorption of photons from the pump pulse, which immediately alters the equilibrium conditions at the moment of excitation, thereby exciting coherent electronic oscillations. This process is analogous to the displacive excitation mechanism of coherent phonons, where the system is excited to a higher electronic state, leading to an impulsive change in the equilibrium position of the atoms and therefore resulting in the coherent excitation of lattice vibrations that maintain the crystal symmetry \cite{Mizoguchi2002}. Because of the indirect nature of the excitation of dopant electrons to the conduction band, the carrier density populated by this mechanism strongly depends on the sample temperature, which explains the fact that the forbidden transition is only observed at elevated temperatures. The theoretical description of this mechanism will be the subject of further research.

Then, we observed changes in different properties of coherent valley-orbit excitation as a function of pump pulse peak intensity, which are shown in Fig. \ref{fig3} for a crystal temperature of $\SI{15}{\kelvin}$. Unfortunately, coherent oscillations vanish quickly for higher peak intensities in the case of the O92 sample with pump pulse polarization along $\left[110\right]$ direction, so results for this case are not present. In the case of the Q8S sample with $\left[110\right]$ orientation and the O92 sample with $\left[100\right]$ orientation, the amplitude of coherent oscillations increases approximately linearly with increasing peak intensity, and above the value of about $\SI{1,5e11}{\watt\per\square\centi\meter}$ the amplitude remains almost constant, creating a plateau. On the other hand, for the Q8S sample with $\left[100\right]$ orientation, the amplitude rises nonlinearly. However, with increasing pump intensity, the growth saturates, indicating the creation of another plateau (Fig. \ref{fig3}a). Also, the ratio between saturated amplitudes in Q8S and O92 samples, both with $\left[100\right]$ orientation, is approximately equal to the ratio between dopant concentrations of the samples. Such behavior thus corresponds to a situation where all phosphorus dopants are coherently excited. Additionally, the decoherence is faster with increasing peak intensity of the pump pulse (Fig. \ref{fig3}b), due to a high density of electrons and holes excited by two-photon absorption driven by the pump pulse. On the other hand, we observe almost constant valley-orbit transition energy (Fig. \ref{fig3}c) and initial phase of the coherent oscillations (Fig. \ref{fig3}d), indicating the situation where $1s\left(A_1\right)\rightarrow1s\left(E\right)$ Raman transition is dominant, what is in agreement with results of temperature dependent measurements below $\SI{30}{\kelvin}$ (Fig. \ref{fig2}c,d).

\begin{figure}[t]
\includegraphics[width=\linewidth]{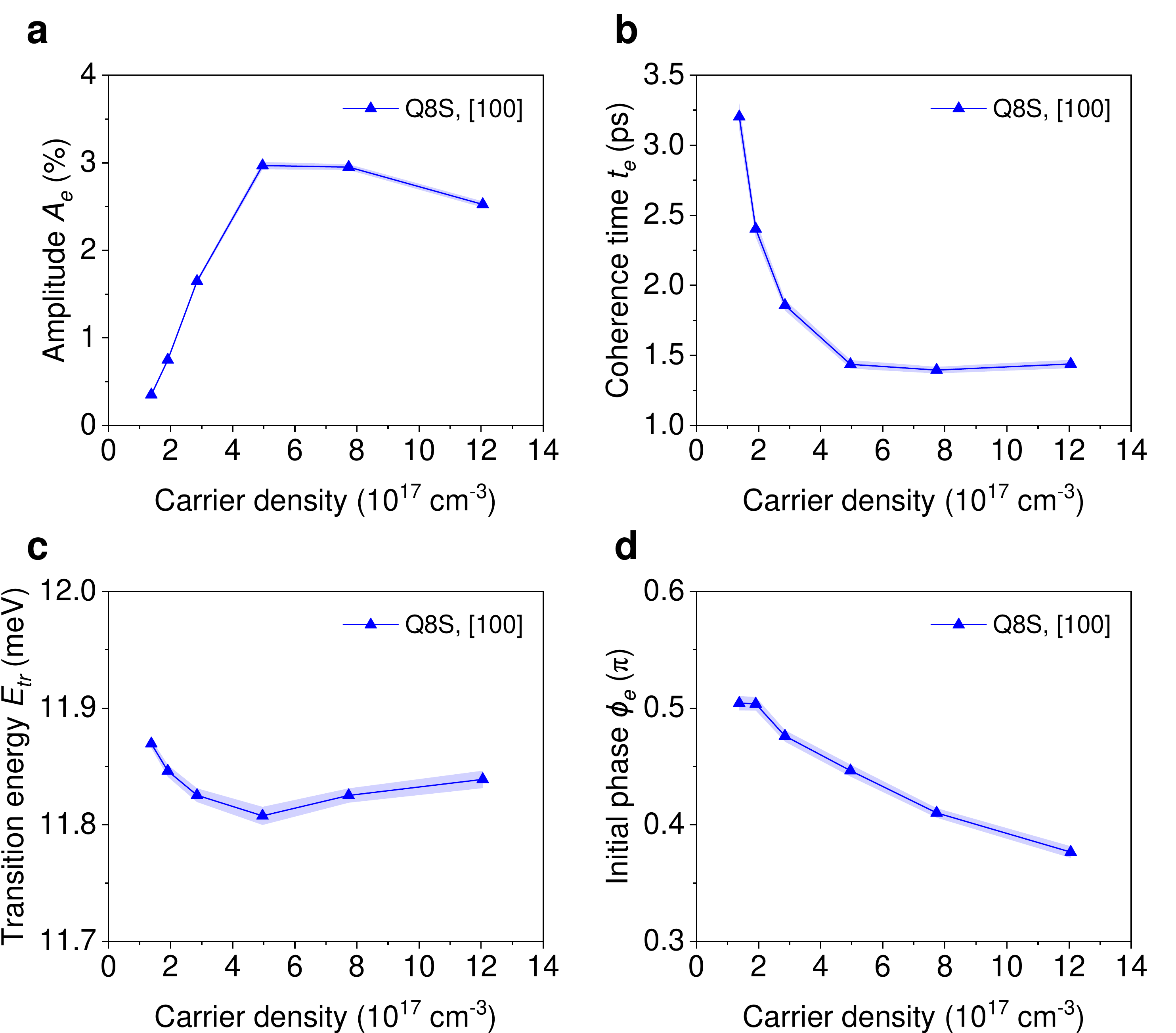}
\centering
\caption{Experimentally measured properties of coherent valley-orbit excitation $1s\left(A_1\right)\rightarrow1s\left(E\right)$ in Q8S at the temperature of 12 K as a function of density of pre-excited carriers. Pump pulse has peak intensity of $I_0 = \SI{1,66e11}{\watt\per\square\centi\meter}$ and polarization oriented along $\left[100\right]$ crystallographic direction. Measured parameters: (\textbf{a}) Amplitude of coherent oscillation, (\textbf{b}) coherence time, (\textbf{c}) central valley-orbit transition energy, and (\textbf{d}) initial phase of coherent oscillations. The shaded areas represent the standard deviation of parameters obtained by fitting experimental data of transient transmittance polarization anisotropy $\Delta T/T_0$ with the function (\ref{eq25:ExpDecayCosExpDecBack}).}
\label{fig4}
\end{figure}

Finally, we added a pre-excitation pulse to our experiment, incident on the sample $\SI{100}{\pico\second}$ before the pump pulse. We studied the influence of Coulomb screening to the properties of coherent valley-orbit excitation as a consequence of high density of pre-excited carriers (Fig. \ref{fig4}) at a temperature of $\SI{12}{\kelvin}$, for the case of the Q8S sample with pump polarization orientation along $\left[100\right]$ direction, where electronic Raman scattering is the strongest. Heavy doping of silicon, as in the Q8S sample, leads to the creation of deep levels within the silicon band gap, including defects such as phosphorus-vacancy complexes (E-centers), di-vacancy of doubly negative charged states \cite{Liu2023}, and carbon-related defects \cite{Gao2018}. These deep-level defects behave as electron traps for bound electrons of phosphorus, leading to the depopulation of phosphorus $1s\left(A_1\right)$ ground level and weakening of the coherent electronic oscillations studied in this work. A pre-excitation pulse excites the trapped electrons to the conduction band, where some of them relax back to dopant levels, leading to the enhancement of the observed Raman transition. This results in the growth of electronic oscillation amplitude up to the threshold value, where all trapped electrons are released, and the amplitude reaches a plateau (Fig. \ref{fig4}a).

The excitation of electrons from deep levels to the conduction band is also reflected in coherence time (Fig. \ref{fig4}b). A stronger Coulomb screening resulting from a higher density of free carriers leads to faster decoherence of the electron wavepacket. Additionally, high dopant concentration results in the formation of a substantial concentration of E-centers \cite{Voronkov2012}. Such phosphorus atoms also contribute to the Raman $1s\left(A_1\right)\rightarrow1s\left(E\right)$ transition, however, their properties would be different than those of phosphorus atoms in a perfect crystal lattice. Furthermore, inhomogeneous distribution of E-centers in silicon lattice could possibly lead to inhomogeneous broadening of $1s\left(A_1\right)\rightarrow1s\left(E\right)$ transition linewidth. This results in shorter coherence time and small changes in transition energy (Fig. \ref{fig4}c). After depopulation of all deep levels, the coherence time remains almost constant. Another hypothesis is that the absorption on the deep levels is comparable to the interband absorption. Then the density of electrons in the conduction band would be limited by the concentration of donors. Accordingly, a more accurate theoretical description of such behavior requires further investigation.

Valley-orbit transition energy (Fig. \ref{fig4}c) and initial phase of oscillations (Fig. \ref{fig4}d) changes only slightly with density of pre-excited carriers, where obtained values are similar to results from measurements without pre-excitation (Figs. \ref{fig2}c,d and \ref{fig3}c,d).

\section{Conclusion}

In conclusion, we present the theoretical and experimental investigation of electronic stimulated Raman scattering by the bound electron transition between valley-orbit split $1s$ states in phosphorus-doped silicon. We studied this phenomenon using transient transmission spectroscopy, where the pump pulse generates a coherent wavepacket as a linear superposition of bound electron eigenstates, and a time-delayed probe pulse investigates its time evolution in the form of oscillations of polarization anisotropy of transient transmittance. This approach offers a possibility to study the bound electron transition between valley-orbit split states from a different point of view compared to previously used methods, including early measurements of photoexcitation spectra \cite{Aggarwal1964, Aggarwal1965, Toyotomi1975} or later measurements of Raman spectra \cite{Jain1976, Wright1967, Stavrias2019}.

The femtosecond temporal resolution, enabled by the ultrashort duration of the utilized laser pulses, allows us to investigate the properties of ultrafast bound electron dynamics, such as the coherence time, valley-orbit transition energy, or the initial phase of coherent oscillations as a function of crystal temperature, peak intensity of the pump pulse, or the density of carriers excited to the conduction and valence bands by a resonant pre-excitation pulse. We observed a decreasing amplitude and coherence time of electronic oscillations at higher temperatures due to incoherent thermal excitation of phosphorus excited state $1s\left(E\right)$, significantly stronger Raman scattering for higher dopant concentrations, as well as faster decoherence of the electron wavepacket as a consequence of the increasing overlap between the wave functions of electrons on neighboring donors. All these observations are consistent with previous measurements \cite{Jain1976, Stavrias2019}.

However, the time resolution allows us to study directly the coherent electronic excitations in the time domain, allowing us to gain better understanding of the excitation mechanisms and to utilize carrier pre-excitation and pump pulse polarization to control the coherent excitation of the valley-orbit-split states. We observed predominant Raman-forbidden displacive impulsive $1s\left(A_1\right)\rightarrow1s\left(T_1\right)$ excitation in sample of phosphorus-doped silicon at temperatures above $\SI{30}{\kelvin}$ with pump pulse linear polarization along $\left[110\right]$ crystallographic direction. We showed that all phosphorus dopants are excited by coherent Raman scattering above a certain threshold peak intensity of the pump pulse, resulting in the saturation of the electronic oscillation amplitude. Finally, the addition of the pre-excitation pulse helps to release electrons from deep-level electron traps created because of the high concentration of dopants in the silicon crystal. Part of the released electrons repopulate the dopant states, leading to an enhancement of the coherent electronic Raman oscillations up to a maximum value, where all deep levels are depopulated by the pump pulse.

The demonstrated coherent time-resolved technique, which combines the well-established method of impulsive stimulated Raman scattering with high temporal resolution of pump-probe measurements, could be used in the future to acquire additional details about electron dynamics introduced by different dopants in silicon or to study the doping of other multivalley semiconductors, which would further enrich the newly emerging field of valleytronics in bulk materials.

\section*{Acknowledgments}
We acknowledge the support given by Charles University (Grant Nos. UNCE/SCI/010, SVV2025–260836, PRIMUS/19/SCI/05 and GA UK 124324), the Czech Science Foundation (Grant No. GA26-21965S) and the European Union (ERC, eWaveShaper, Grant No. 101039339). This work was supported by TERAFIT (Project No. CZ.02.01.01/00/22\_008/0004594), which was funded by OP JAK, call Excellent Research. Z. \v{S}ob\'{a}\v{n} acknowledges the support of MEYS grant LM2023051.

\section*{Data availability}

The data that support the findings of this article are openly available \cite{Zenodo}.

\nocite{*}

\bibliography{articlereferences}

\end{document}